\documentclass[aps,preprint,amsmath,preprintnumbers]{revtex4-1}

\usepackage[colorlinks=true
,urlcolor=blue
,anchorcolor=blue
,citecolor=blue
,filecolor=blue
,linkcolor=blue
,menucolor=blue
,pagecolor=blue
]{hyperref}
\usepackage[all]{hypcap}

\newcommand{\beq}{\begin{equation}}
\newcommand{\eeq}{\end{equation}}
\renewcommand{\slash}[1]{#1\!\!\!/}
\newcommand{\slashvec}[1]{\vec{#1}\!\!\!/}

\newcommand{\MET}{{\slash E_T}}
\newcommand{\METvec}{{\slashvec E_T}}
\newcommand{\pol}{{\cal P}}

\begin{document}

\title{Measuring polarization of light quarks at ATLAS and CMS}

\author{Yevgeny Kats}
\email{yevgeny.kats@weizmann.ac.il}
\affiliation{Department of Particle Physics and Astrophysics\\
             Weizmann Institute of Science\\
             Rehovot 7610001, Israel}

\begin{abstract}
Polarization of strange quarks is preserved to a high degree when they hadronize into $\Lambda$ baryons, as observed in $Z$ decays at LEP. This opens up the possibility for ATLAS and CMS to use strange-quark polarization measurements as a characterization tool for new physics scenarios that produce such quarks. Measurements in $t\bar t$ samples would be useful for obtaining additional information about the polarization transfer from the strange quark to the $\Lambda$ baryon. Already with 100~fb$^{-1}$ in Run~2, $t\bar t$ samples in ATLAS and CMS become competitive in sensitivity with the $Z$ samples of the LEP experiments. Moreover, while the LEP measurements were done inclusively over all quark flavors, which makes their interpretation dependent on various modeling assumptions, $t\bar t$ events at the LHC offer multiple handles for disentangling the different contributions experimentally. We also discuss the possibility of measuring polarizations of up and down quarks.
\end{abstract}

\maketitle

\section{Introduction and Overview}

While it is easy for the LHC detectors to reconstruct the momentum of an energetic quark by measuring the jet it produces, there is no straightforward way to determine the quark's polarization. In general, three obstacles are present. First, the nonperturbative strong interactions during the QCD-scale fragmentation process can change the polarization of the quark or perhaps even wash it out completely. Second, even if part of the polarization survives the QCD-scale interactions, it is usually impossible to know which of the many hadrons in the jet carries the original quark. Third, the polarization can disappear during the hadron's lifetime and intermediate decays, before the electroweak decay that can actually provide information about the polarization.

For a heavy quark, $m_q \gg \Lambda_{\rm QCD}$, such as the bottom (and to a large extent also the charm), the situation is simpler~\cite{Mannel:1991bs,Ball:1992fw,Falk:1993rf,Galanti:2015pqa}. First, since the quark's chromomagnetic moment is $\propto 1/m_q$, QCD-scale interactions cannot change its polarization significantly. Second, the hadron containing the original quark is typically the most energetic one. The energy fraction carried by the other hadrons is expected to be only ${\cal O}(\Lambda_{\rm QCD}/m_q)$ at low energy scales~\cite{Bjorken:1977md,Peterson:1982ak,Neubert:2007je}. At the electroweak scale, this number is measured to be around 30\% in the $b$-quark case~\cite{Heister:2001jg,DELPHI:2011aa,Abbiendi:2002vt,Abe:2002iq} and 50\% in the $c$-quark case~\cite{Barate:1999bg}, consistent with renormalization group evolution~\cite{Cacciari:2005uk}. The third difficulty is largely intact. In most cases the quark ends up in a meson, which decays down to a ground-state meson of the corresponding heavy flavor prior to the electroweak decay of the quark itself. Since the ground-state mesons are scalars (and the intermediate decays are of very limited use~\cite{Falk:1993rf}), the polarization information is lost. However, the story is different if the quark hadronizes into a baryon. Since baryons have spin, they can carry information about the initial polarization. In the $\Lambda_b$ baryon, in particular, the light degrees of freedom are in a spin-0 state (in the heavy quark limit), so the $b$ spin is not subject to any interactions throughout the $\Lambda_b$ lifetime~\cite{Mannel:1991bs,Ball:1992fw,Falk:1993rf}. In other baryons, such as the $\Sigma_b$ and $\Sigma_b^\ast$ whose light degrees of freedom are spin-1, the $b$ polarization does change. The $\Sigma_b^{(\ast)}$'s then decay to the $\Lambda_b$, plus a soft pion, reducing the polarization of the inclusive $\Lambda_b$ sample. But at the end of the day, the $\Lambda_b$ polarization simply gets reduced by a fixed ${\cal O}(1)$ factor relative to the original $b$-quark polarization~\cite{Falk:1993rf,Galanti:2015pqa}, allowing for measurements of the latter. Evidence of this effect has been observed at LEP~\cite{Buskulic:1995mf,Abbiendi:1998uz,Abreu:1999gf}. The precise polarization retention factor can be measured in single-lepton $t\bar t$ samples at the LHC and then used in $b$-quark polarization measurements in new physics samples~\cite{Galanti:2015pqa}.

The case of the strange quark, whose mass is of order $\Lambda_{\rm QCD}$, na\"{i}vely does not look promising. There is no guarantee that any polarization survives the QCD-scale fragmentation process. Furthermore, the hadron carrying the original strange quark is not necessarily the most energetic one, and, unlike for heavy quarks, secondary strange hadrons are frequently produced in fragmentation and dilute the inclusive polarization.

Nonetheless, and despite the lack of good theoretical control over the processes that affect the polarization of strange quarks during fragmentation, we argue that reliable measurements of longitudinal polarization of strange quarks by ATLAS and CMS, using $\Lambda$ baryons, are feasible. It is known experimentally from measurements in $Z$ decays at LEP~\cite{Buskulic:1996vb,ALEPH:1997an,Ackerstaff:1997nh}, as we review, that the $\Lambda$'s in fact preserve an ${\cal O}(1)$ fraction of the original longitudinal polarization of the strange quark. Thanks to factorization of short and long-distance physics, the fraction measured at LEP will apply to any measurement in Standard Model or new physics processes by ATLAS or CMS, as long as the energy scale of the process is similar to the $Z$ boson scale. For processes with a very different energy scale, one only needs to add the effects of renormalization group evolution of fragmentation functions, which are under good theoretical control~\cite{deFlorian:1997zj}. The polarization measurement can be validated, and higher precision on the spin-dependent fragmentation functions obtained, using Standard Model $t\bar t$ events, where polarized strange quarks are available from $W^+ \to c\bar s$ decays.

Eventually, measurements of $u$ and $d$ quark polarizations might become possible as well. While the proton and neutron do not decay within the detectors, one may use the less common events in which the $u$ or $d$ produce a $\Lambda$. While in the na\"ive quark model the spin of the $\Lambda$ is carried entirely by the $s$ quark, data on nucleon spin structure suggest that a sizable part of the spin in the $\Lambda$ is actually carried by the $u$ and $d\,$~\cite{Burkardt:1993zh}. This may imply that $\Lambda$'s produced in the fragmentation of $u$ and $d$ quarks preserve some fraction of the quark's polarization~\cite{Burkardt:1993zh,Jaffe:1996wp,Kotzinian:1997vd,Boros:1998kc}. Furthermore, even in the na\"ive picture, the $u$ and $d$ quarks do carry spin in heavier baryons such as the $\Sigma$ and $\Sigma^\ast$. When these baryons are produced from a $u$ or $d$ quark and then decay to a $\Lambda$, some of the quark's initial polarization is expected to be preserved in the $\Lambda$ polarization~\cite{Bigi:1977qe,Kotzinian:1997vd}. Polarized lepton or neutrino deep inelastic scattering experiments~\cite{Alekseev:2009ab,Airapetian:2006ee,Adams:1999px,Astier:2000ax} and polarized $pp$ collisions~\cite{Abelev:2009xg,Deng:2014eha} can in principle provide useful information on this question. However, they are currently limited by statistics and uncertainties regarding the strange parton distribution functions~\cite{Alekseev:2009ab,Deng:2014eha}. ATLAS and CMS can attempt to observe the effect more directly using $W^+\to u\bar d$ decays in $t\bar t$ samples.

We note in passing that in principle there is no need in an electroweakly decaying baryon, like the $\Lambda$, to be sensitive to the quark polarization. Certain observables constructed from momenta of two or more hadrons in the jet (``jet handedness'') can have polarization-dependent distributions due to QCD processes alone~\cite{Nachtmann:1977ek,Efremov:1978qy,Efremov:1992pe}. However, the size of this effect is unknown, and a significant upper bound on it was obtained by the SLD experiment~\cite{Abe:1994bk,Abe:1995wh}.

\section{\texorpdfstring
{LEP measurements of $\Lambda$ polarization in $Z$ decays}
{LEP measurements of Lambda polarization in Z decays}}

The longitudinal polarization of strange quarks produced via $e^+e^- \to Z \to s\bar s$, including the one-loop QCD effects~\cite{Korner:1993dy}, is
\beq
\pol(s) \simeq -0.91 \,.
\eeq
The LEP experiments ALEPH~\cite{Buskulic:1996vb,ALEPH:1997an} and OPAL~\cite{Ackerstaff:1997nh} measured the polarization of $\Lambda$'s in inclusive samples of $e^+e^- \to Z \to jj$ events. To reduce the contribution from the mostly soft $\Lambda$'s produced from secondary strange quarks, it is useful to apply a lower cut on the energy (or momentum) fraction $z$ carried by the $\Lambda$ relative to the whole jet. For $z > 0.4$, the experiments report
\begin{align}
&\pol(\Lambda) = -0.41 \pm 0.07 \qquad\mbox{ALEPH~\cite{ALEPH:1997an}}\,,\\
&\pol(\Lambda) = -0.46 \pm 0.12 \qquad\mbox{OPAL~\cite{Ackerstaff:1997nh}}\,,
\end{align}
and for $z > 0.3$
\begin{align}
&\pol(\Lambda) = -0.31 \pm 0.05 \qquad\mbox{ALEPH~\cite{ALEPH:1997an}}\,,\\
&\pol(\Lambda) = -0.33 \pm 0.08 \qquad\mbox{OPAL~\cite{Ackerstaff:1997nh}}\,.
\end{align}
These results demonstrate that a large fraction of the original $s$-quark polarization is retained in the $\Lambda$'s. In fact, as simulated by ALEPH and OPAL following the approach of~\cite{Gustafson:1992iq} and using {\sc Jetset}~\cite{Sjostrand:1993yb}, these numbers can be reproduced, within large modeling uncertainties, even if one assumes that there is no polarization loss at all for $\Lambda$'s carrying the primary strange quarks. Instead, $\pol(\Lambda)$ gets diluted by the partly polarized $\Lambda$'s produced in decays of other primary strange baryons (most importantly $\Sigma^0\to\Lambda\,\gamma$, $\Sigma^{\ast\pm,0} \to \Lambda\,\pi^{\pm,0}$ and $\Xi^{0,-} \to \Lambda\,\pi^{0,-}$) and unpolarized $\Lambda$'s from nonprimary strange quarks or from decays of heavy-flavor mesons, and accounts also for $\Lambda$'s from the parity-violating (electroweak) decays of heavy-flavor baryons and nonprimary $\Xi$'s whose contributions are slightly polarized due to a selection bias.

We note from examining the simulation results of ALEPH and OPAL that for a sample containing only strange jets one would have $-0.65 \lesssim \pol(\Lambda) \lesssim -0.49$ for $z > 0.3$ and $-0.69 \lesssim \pol(\Lambda) \lesssim -0.53$ for $z > 0.4$, where the precise numbers are somewhat uncertain due to modeling assumptions. The depolarization relative to $\pol(s)$ would be primarily due to $\Lambda$ production from the decays of heavier strange baryons, which is comparable to direct $\Lambda$ production, and to a lesser extent due to $\Lambda$'s with secondary strange quarks.

The possibility that $\Lambda$'s carrying primary $u$ and $d$ quarks preserve a fraction of the quark polarization (with an opposite sign)~\cite{Burkardt:1993zh} was examined by ALEPH~\cite{ALEPH:1997an} (see also~\cite{Kotzinian:1997vd,deFlorian:1997zj,Boros:1998kc}) and found to be allowed by their (flavor-inclusive) data. As we will see, $t\bar t$ samples at the LHC provide a great opportunity to explore this question in more detail by measuring the different quark flavors separately.

\section{\texorpdfstring
{Measuring $\Lambda$'s at ATLAS and CMS}
{Measuring Lambdas at ATLAS and CMS}}
Polarization of strange (and possibly also up and down) quarks can be accessed by ATLAS and CMS by measuring the polarization of $\Lambda$ baryons in the jets. The most common decay of the $\Lambda$, with a branching fraction of 64\%~\cite{PDG}, is
\beq
\Lambda \to p\,\pi^- \,,
\label{decay}
\eeq
giving rise to an easily identifiable signature of a pair of oppositely charged tracks that originate from a highly displaced vertex ($c\tau_\Lambda \simeq 7.9$~cm) and reconstruct the $\Lambda$ mass. The background from $K_S^0 \to \pi^+\pi^-$ decays (which are also displaced, $c\tau_{K_S^0} \simeq 2.7$~cm) can be largely eliminated by requiring that the tracks do not reconstruct the $K_S^0$ mass if both are assigned the $\pi^\pm$ mass. The polarization can be determined from the angular distribution of the decay,
\beq
\frac{d\,\Gamma}{d \cos \theta} \;\propto\; 1 + \alpha\,\pol(\Lambda)\cos\theta
\eeq
where $\theta$ denotes, in the $\Lambda$ rest frame, the direction of motion of the proton relative to the polarization direction, and~\cite{PDG}
\beq
\alpha = 0.642 \pm 0.013 \,.
\label{alpha}
\eeq
The same distribution describes the decays of the antiparticle, $\overline\Lambda$, of an opposite polarization. The same decay channel was used at LEP. The only other significant decay, $\Lambda \to n\,\pi^0$, is more difficult to identify.

In both ATLAS and CMS, several analyses have already reconstructed $\Lambda \to p\,\pi^-$ decays, either for studies of the $\Lambda$ itself~\cite{Khachatryan:2011tm,Aad:2011hd,ATLAS:2014ona} (including a measurement of the transverse polarization in inclusive production~\cite{ATLAS:2014ona}) or for studies of the $\Lambda_b$ using the decays $\Lambda_b\to J/\psi(\to\mu^+\mu^-)\,\Lambda$~\cite{ATLAS:2011uha,Chatrchyan:2012xg,Aad:2012bpa,Chatrchyan:2013sxa,CERN-THESIS-2013-218,Aad:2014iba}. For $\Lambda$'s with $p_T \sim 5$~GeV, CMS reported an efficiency of around $15\%$ in~\cite{Khachatryan:2011tm}, and ATLAS $35\%$ in~\cite{Aad:2011hd} and roughly $20\%$ in~\cite{ATLAS:2014ona}; the differences are primarily due to different selection requirements. The background under the mass peak is quite small in all cases. For new physics samples, much higher values of $p_T$ will likely be relevant. The efficiency will then possibly be significantly lower because track reconstruction becomes more difficult for larger displacements. In particular, the $\Lambda$ efficiency in ATLAS~\cite{Aad:2011hd} for transverse flight distances between 10 and 30~cm is only 10\%, and becomes yet lower for larger distances. In the LEP measurements in $Z$ decays, the efficiencies were in the same ballpark --- varying from 36\% for $z = 0.15$ to 13\% for $z = 0.5$ in ALEPH~\cite{Buskulic:1996vb}, and from 25\% for $z = 0.15$ to 10\% for $z = 0.4$ in OPAL~\cite{Ackerstaff:1997nh}. In the future, the situation in ATLAS and CMS will likely improve significantly with the planned upgrades to their tracking detectors~\cite{CERN-LHCC-2012-022,CMS:2012sda,CERN-LHCC-2011-006}.

Before applying the strange-quark polarization measurement to new physics samples, it can be validated on Standard Model samples of polarized strange quarks. Such measurements would also be useful for gaining additional information about the relevant fragmentation functions, which describe the abundance and polarization of $\Lambda$'s in jets as a function of their energy fraction $z$. Good knowledge of the fragmentation functions will allow computing the scale dependence of the polarization transfer effect, as well as provide further input to the theoretical understanding and modeling of fragmentation (see, e.g.,~\cite{deFlorian:1997zj}).

At the LHC, the hadronic decays of the $Z$ suffer from a very large QCD background, which would make the measurement more difficult than at LEP. However, the LHC experiments have the opportunity to use clean $t\bar t$ samples, where longitudinally polarized strange quarks are produced in $W^+ \to c\bar s$ decays. After accounting for the one-loop QCD effects~\cite{Korner:1993dy}, the $s$-quark polarization in $W$ decays is $\pol(s) \simeq -0.97$.

\section{\texorpdfstring
{Proposed studies in $t\bar t$ samples}
{Proposed studies in ttbar samples}}

Let us consider a measurement in the Run~2 data, which we assume will be based on 100~fb$^{-1}$ of 13~TeV collisions. The $t\bar t$ production cross section is $\sigma \approx 800$~pb~\cite{Kidonakis:2014isa,Muselli:2015kba}. Accounting for the branching fractions for $W \to cs$ on one side of the event and $W \to \ell\nu$ (with $\ell = e$ or $\mu$) on the other (to reduce backgrounds), and assuming this final state will have an efficiency of about 20\% in a standard single-lepton~+ jets $t\bar t$ selection (e.g.,~\cite{Aad:2015pga}), we obtain a sample containing roughly $2.2 \times 10^6$ polarized strange quarks (including antiquarks). Thus, already in Run~2 the statistics will be comparable to LEP, where each experiment had about $1.9 \times 10^6$ slightly less polarized strange quarks (including antiquarks) from $Z$ decays~\cite{Ackerstaff:1997nh,ALEPH:1997an}.

A standard kinematic reconstruction of single-lepton~+ jets $t\bar t$ events determines which jets originate from the $W$ decay, but not their flavor. However, since energetic $\Lambda$'s are produced much more frequently in $s$-initiated jets than in other jets, it would not be unreasonable to even collect $\Lambda$'s from all jets that originate from a $W$. Recall that the measurements at LEP were also done inclusively over comparable numbers of $u$, $d$, $s$ and $c$ jets, and in addition $b$ jets. Still, for easier interpretation of the results, and more straightforward application to new physics scenarios, which may contain only strange quarks, or a different mixture of several quark flavors, it would be useful to suppress contributions from jets that do not originate from strange quarks. To improve the sensitivity to the polarization, it would also be useful to reduce the contributions from secondary strange quarks, which produce mostly unpolarized $\Lambda$'s, in all jets.

One handle readily available in single-lepton $t\bar t$ samples is the charge of the lepton, which determines whether to expect a $\Lambda$ or a $\overline\Lambda$ on the other side of the event. Without any cost in the polarized signal efficiency, this reduces the contribution from secondary $\Lambda$'s by a factor of~2 and entirely eliminates the contribution from primary $\Lambda_c$ decays in $c$ jets.

Another handle is charm tagging, which can be used for distinguishing between the $W\to cs$ and $W\to ud$ decays, and between the $c$ and $s$ jets in $W\to cs$. It also reduces the number of ways in which a radiation jet with $g\to s\bar s$ can contribute to the sample by being misidentified in the kinematic reconstruction as one of the jets from the hadronic $W$ decay. ATLAS has already developed a charm tagging algorithm and used it in Run~1~\cite{ATL-PHYS-PUB-2015-001,Aad:2014nra,Aad:2015gna}. To select a clean sample of $s$-initiated jets, one can tag the accompanying $c$ jets using, for example, the operating point with $\epsilon_c \approx 0.20$ and $\epsilon_{udsg} \approx 0.005$. The $b$-jet rejection power of the operating point is not important to us since $b$'s are not produced in $W$ decays. In Run~2, the charm tagging performance is actually expected to be even better than these numbers thanks to the ``Insertable B-Layer'' of pixels that has been added to the ATLAS detector~\cite{ATL-PHYS-PUB-2015-001,ATLAS-TDR-19,Aad:2012wf}.

Besides the measurement in strange jets, measurements of the $\Lambda$ polarization in $c$ jets (from $W \to cs$), $u$ and $d$ jets (from $W \to ud$) and $b$ jets (from $t\to Wb$) would also be very beneficial. They would facilitate the comparison with the LEP results and provide useful input for possible future measurements in new physics scenarios in which strange and other quark flavors appear together. Most interestingly, isolating polarized $\Lambda$ contributions from $u$ and $d$ jets would provide a unique input to theory, and open the way for measurements of $u$ and $d$ quark polarizations in general settings.

Care must be taken when defining a sample of $c$ jets using a charm tagging algorithm. A large fraction of $\Lambda$'s in $c$ jets originates from $\Lambda_c$ decays. Since the $\Lambda_c$ is not as long lived as the $D$ mesons ($\tau_{\Lambda_c} \approx \tau_{D^0}/2 \approx \tau_{D_s^\pm} / 2.5 \approx \tau_{D^\pm} / 5$~\cite{PDG}), the tagging efficiency for events with a $\Lambda_c$ might be lower than the inclusive charm tagging efficiency.

Samples enriched in $u$ and $d$ jets can be obtained by vetoing on a loose charm tag (e.g., defined using the operating point with $\epsilon_c \approx 0.50$, $\epsilon_{udsg} \approx 0.18$ from~\cite{ATL-PHYS-PUB-2015-001}). Due to the isospin symmetry, $u$ and $d$ jets are expected to have approximately the same properties. Still, it may be useful to separate between them for better control of the large backgrounds due to $W\to cs$ decays that fail charm tagging. One can use the charge of the lepton on the other side of the event to determine which of the two, the $u$ or the $d$ jet, is expected to contain a $\Lambda$ and which a $\overline\Lambda$ as a primary hadron. Then the polarized background for $u$ jets will be coming from $c$ jets, and for $d$ jets from $s$ jets.

Let us end this section with an estimate of the expected sensitivity of the strange-quark polarization measurement in $t\bar t$ in Run~2. OPAL~\cite{Ackerstaff:1997nh} reports the number of $\Lambda$'s (including $\overline\Lambda$'s) in their $Z$ sample, after reconstruction and background subtraction, to be approximately 8309 for $z > 0.3$ and 2950 for $z > 0.4$. Based on OPAL's simulation, in both cases roughly half of these $\Lambda$'s come from strange-initiated jets. Strange jets from $t\bar t$ events will have approximately the same properties. Corrections due to renormalization group evolution are expected to be small because the energies of the strange quarks from $Z$ production at LEP and $t\bar t$ production at LHC are comparable. The $\Lambda$ reconstruction efficiencies in ATLAS and CMS are roughly the same as in OPAL, as we discussed. Then, with the event yield estimated above, and including the charm tagging efficiency, the polarization measurement samples will consist of roughly 1000 strange jets with $\Lambda$'s with $z > 0.3$. Taking into account that $\pol(s)$ is slightly larger in the $t\bar t$ case than in the $Z$ case, the $\Lambda$ polarization is expected to be $\pol(\Lambda) \approx -0.6$. The achievable statistical precision of the polarization measurement, taking into account the spin analyzing power from Eq.~\eqref{alpha}, is then roughly 16\%~\footnote{Here and in the following, we estimate the statistical uncertainty as in Sec.~4.2 of~\cite{Galanti:2015pqa}. Backgrounds due to $K_S^0 \to \pi^+\pi^-$ and random track combinations, which depend on the experimental details but are expected to be small, are neglected. ATLAS and CMS results are not combined.} already in Run~2 of the LHC.

\section{New physics example}

The strange-quark polarization measurement can be used in any new physics scenario in which such quarks are produced. It will be most effective in cases where the current uncertainty on the background estimate in the relevant searches is large, since that can allow the signal to be sufficiently large so that enough statistics can be collected for the measurement after the signal is discovered. The statistics needs to be large because the probability for a $\Lambda$ (with $z > 0.3$, for example) to be present in a strange-initiated jet is only ${\cal O}(3\%)$~\cite{Albino:2008fy} and the reconstruction efficiency for high-$p_T$ $\Lambda$'s is only ${\cal O}(10\%)$. Let us present a simple example of a scenario in which the measurement would be useful.

Suppose an excess is observed in searches for jets and missing energy, and it is being suggested, based on the size and kinematic properties of the excess, that it is due to the pair production of the right-handed strange squark, decaying to a strange quark and a stable bino:
\beq
pp \to \tilde s_R\tilde s_R^\ast\,,\quad
\tilde s_R \to s\,\tilde\chi_1^0 \,.
\label{NP}
\eeq
(The same excess cannot be attributed to $\tilde s_L$, because $\tilde s_L$ forms a doublet with $\tilde c_L$, which would have been observed even earlier in searches that use charm tagging.) A possible test of this interpretation would be to measure the strange-quark polarization, which is expected to be right handed, i.e.\ $\pol(s) \approx +1$, since the bino coupling does not mix chiralities.

Most sensitive to this scenario are the ATLAS and CMS searches for jets (or a monojet) and missing energy~\cite{Aad:2014nra,Aad:2014wea,CMS:2014yma,Chatrchyan:2014lfa}. Consider the case of $m_{\tilde s_R} = 200$~GeV, $m_{\tilde\chi_1^0} = 150$~GeV. Neither the search~\cite{Aad:2014nra} (see their Fig.~9, left), nor~\cite{Aad:2014wea} (see their Fig.~10c) nor~\cite{CMS:2014yma} (see their Fig.~7, right) excludes this scenario, while the search~\cite{Chatrchyan:2014lfa} (their Fig.~7a) does not present the relevant range of masses. At the same time, the scenario is close to the regions excluded by these searches, so it will likely be possible to discover it in Run~2.

To estimate the polarization measurement sensitivity for this scenario, let us consider the most sensitive search region of~\cite{Aad:2014wea} for the event selection, even though it was optimized for the 8~TeV search rather than the future polarization measurement. It is the 2~jets+$\MET$ loose search region, 2jl~\footnote{Region 2jl of~\cite{Aad:2014wea} is defined by $\MET > 160$~GeV, $p_T(j_{1,2}) > 130,60$~GeV, $\Delta\phi(j_{1,2,(3)},\METvec) > 0.4$, $\MET/\sqrt{H_T} > 8$~GeV$^{1/2}$, $m_{\rm eff}({\rm incl.}) > 800$~GeV, lepton veto.}, in which $13000 \pm 1000$ background events and about 1300 signal events~\footnote{The signal acceptance times reconstruction efficiency is available in Fig.~15b in the auxiliary material for~\cite{Aad:2014wea}.} were expected. For 3~ab$^{-1}$ of data at 14~TeV, with the crude approximation that the selection efficiencies remain the same~\footnote{While the general expectation, that trigger requirements will become harsher with the increase in energy and luminosity, suggests the efficiencies will decrease, it is also reasonable to believe that once new physics is discovered the triggers will be adjusted in its favor.} and only the cross sections get rescaled (by a factor of $4.2$ for the signal~\cite{Borschensky:2014cia} and roughly $2$ for the background~\cite{Mangano:2012mh}, which is mostly due to $Z/\gamma^*$+jets and $W$+jets), we obtain $8.1 \times 10^5$ signal events and $3.8 \times 10^6$ background events.

The signal events will contain jets due to both the $s$ (and $\bar s$) quarks from Eq.~\eqref{NP} and initial-state radiation (ISR). It is a question for a detailed simulation to optimize the requirement for a jet to be included in the polarization measurement (e.g., the 3 leading jets, or all jets with $p_T > 30$~GeV, or a more sophisticated criterion). Note that since only about $1/3$ of the ISR jets will contain moderately energetic strange quarks (via $g\to s\bar s$), the contribution of an ISR jet to a sample of high-$z$ $\Lambda$'s will be suppressed, on average, relative to that of a strange jet. In the dominant backgrounds, $Z(\to\nu\nu)$+jets and $W(\to \tau_h\nu)$+jets, the jets are mostly ISR jets, along with a small number of hadronic tau jets, so their contributions to the sample will be suppressed as well. Since the $s$ and $\bar s$ jets from Eq.~\eqref{NP} are relatively soft, suppose that, on average, only one such jet per event will be included in the sample. With the assumption justified earlier, that the probability for a strange jet to contain a reconstructed $\Lambda$ with $z > 0.3$ is similar to OPAL, the sample will contain about 1800 $\Lambda$'s and $\overline\Lambda$'s from the signal. Background includes jets from the background processes (suppose 3 per event) as well as ISR jets from the signal (suppose 2 per event). The abundance of $\Lambda$'s (including $\overline\Lambda$'s) with $z > 0.3$ in gluon jets is lower than in $s$ jets by a factor of about~3.3~\cite{Albino:2008fy}. Then, for the purpose of a rough estimate taking all background jets to be gluon jets, the signal fraction in the sample of $\Lambda$'s will be about $17\%$. This would allow us to determine the $s$-quark polarization with statistical precision of $30\%$. If the $\Lambda$ reconstruction efficiency with upgraded ATLAS and CMS tracking detectors ends up being a factor of $5$ higher than the ${\cal O}(10\%)$ efficiency of OPAL, the precision becomes 13\%. The sensitivity could be further improved by optimizing the event selection and the cut on $z$ and combining ATLAS and CMS results. This will be especially important for more statistically challenged new physics scenarios.

\acknowledgments

I am thankful to Diptimoy~Ghosh, Andrea~Giammanco, Yuval~Grossman, Yotam~Soreq and Emmanuel~Stamou for valuable comments.

\bibliographystyle{utphys}
\bibliography{s_pol}

\end{document}